\documentclass[preprint,showpacs,preprintnumbers,amsmath,amssymb,endfloats*]{revtex4}
\usepackage{graphicx}

\begin{document}

\thispagestyle{empty}

\title{Possibility of measuring the thermal Casimir interaction
between a plate and a cylinder attached to a micromachined oscillator}

\author{R.~S.~Decca,${}^1$ E.~Fischbach,${}^2$
G.~L.~Klimchitskaya,${}^3$
D.~E.~Krause,${}^{4,2}$ D.~L\'{o}pez,${}^5$
and V.~M.~Mostepanenko${}^6$
}

\affiliation{
${}^1$Department of Physics, Indiana University-Purdue
University Indianapolis, Indianapolis, Indiana 46202, USA\\
${}^2$Department of Physics, Purdue University, West Lafayette, Indiana
47907, USA\\
${}^3$North-West Technical University, Millionnaya Street 5, St.Petersburg,
191065, Russia \\
${}^4$Physics Department, Wabash College, Crawfordsville, Indiana 47933,
USA\\
${}^5$Center for Nanoscale Materials, Argonne National Laboratory,
Argonne, Illinois 60439, USA \\
${}^6$Noncommercial Partnership
``Scientific Instruments'',  Tverskaya Street 11, Moscow,  103905, Russia
}

\begin{abstract}
We investigate the possibility of measuring the thermal Casimir force
and its gradient in the configuration of a plate and a
microfabricated cylinder attached to a
micromachined oscillator. The Lifshitz-type
formulas in this configuration are derived using the proximity
force approximation. The accuracy for the obtained expressions is
determined from a comparison with exact results available
in ideal metal case. Computations of
the thermal correction to both the Casimir force
and its gradient are performed in the framework of different theoretical
approaches proposed in the literature. The correction to the Casimir force
and its gradient due to lack of parallelism of the plate and
cylinder is
determined using the nonmultiplicative approach. The error introduced
in the theory due to the finite length  of the cylinder is estimated.
We propose that both static and dynamic experiments measuring
the thermal
Casimir interaction between a cylinder and a plate using a
micromachined oscillator can shed additional light on the
 thermal Casimir force problem. Specifically, it is shown that the
 static experiment is better adapted for the measurement of
 thermal effects.
\end{abstract}

\maketitle

\section{Introduction}

The Casimir effect which was predicted more than 60 years ago
\cite{1} has recently attracted much theoretical and experimental
attention (see the monographs \cite{2,3,4,5,6} for a overview of
the subject). The Casimir force originates from quantum
fluctuations of the electromagnetic field. Thus, it is of the
same origin as the van der Waals force, but acts at larger
separations between the interacting surfaces where
relativistic retardation effects become significant.
The unified theory of both the van der Waals and Casimir forces
was developed by Lifshitz \cite{7,8}. It describes the free energy
and force of the van der Waals and Casimir interaction using the
frequency-dependent dielectric permittivity of interacting
bodies.

The Casimir force acting between two neutral surfaces becomes
dominant at separations below $1\,\mu$m. Until the last decade
there was a lack of experimental information about this
quantum phenomenon, but this situation has
changed in recent years due to rapid progress in
nanotechnology. Following two landmark experiments
\cite{9,10}, many measurements of the Casimir and
Casimir-Polder (atom-wall) forces have been performed with metal,
semiconductor and dielectric surfaces (a review of all experiments
can be found in Ref.~\cite{11}).

The experimental and theoretical investigation of the Casimir
force at nonzero temperature produced an unexpected result.
It turned out that if conduction electrons in metal plates with
perfect crystal lattices are described by the dissipative Drude
model, this results in a violation of the Nernst heat theorem
in the Lifshitz theory \cite{6,11,12,13}. The same takes place
for dielectric-type semiconductor and dielectric test bodies if
free charge carriers, which are unavoidably present at nonzero
temperature, are described by the Drude-type contribution to
the dielectric permittivity \cite{6,11,14,15,16}.
The parameter range, where a violation occurs, was recently
rederived and extended also to the geometry of a sphere
above a plate \cite{16A}.
It was
demonstrated that for metals with impurities the Nernst heat
theorem in the Lifshitz theory is preserved \cite{17,17a,18,19}.
However, in the case of metals with perfect crystal structure
(which is the basic model used in condensed matter physics),
and for dielectrics, a satisfactory solution was not found.
This even resulted in the attempts \cite{20,21} to modify
the Lifshitz theory, and this led to a controversial
discussion in the literature \cite{6,11,22,23,24,25,26,27,28,29}.
{}From the experimental side, it was shown that data exclude
the Lifshitz theory, if the role of conduction
electrons is taken into account
by means of the dissipative Drude model. This was
repeatedly demonstrated for metal \cite{30,31,32,33,34},
semiconductor \cite{35,36} and dielectric \cite{37,38}
test bodies.

All modern experiments measuring the Casimir force (with the
single exception of the experiment of Ref.~\cite{39} which was
burdened by rather large error of about 15\%) were performed
using the configuration of a sphere above a plate. Bearing in
mind that contradictions between Lifshitz theory and
thermodynamics, and experimental data discussed above,
are of great concern for the foundations of quantum statistical
physics \cite{40}, it is worthwhile to consider the possibilities
of alternative experiments. One more configuration of great
promise that was discussed in the literature \cite{41A,41} is a
cylinder parallel to a plane plate. In some sense it is
intermediate between the configurations of two parallel plates
and a sphere above a plate, by preserving some advantages of the
latter while making the problem of preserving the parallelism
less difficult than for two plates. An additional advantage is that the
Casimir force in the cylinder-plate geometry was recently calculated
exactly in an ideal metal case
\cite{42,43,44A}. This
allows one to reliably estimate possible errors introduced from the
use of the proximity force approximation.

The original proposal \cite{41} considered a relatively large
cylinder with length $L=2\,$cm and diameter $2R=6.35\,$mm
separated from the plate with a wide gap $a>1\,\mu$m.
It has been shown \cite{44,45}, however, that for metal
coated spherical lenses with centimeter-size curvature radii,
the electrostatic calibration of Casimir apparatus meets
serious difficulties. Specifically, the residual potential
difference may become dependent on separation, and the
force-distance relation for the electric force deviates from the
form predicted by classical electrodynamics in a sphere-plane
geometry. (This does not happen in the experiments of
Refs.~\cite{30,31,32,33,34} dealing with spheres of about
$150\,\mu$m in radius interacting with the plate of an
oscillator.) Similar anomalies were reported in Ref.~\cite{46}
for the configuration of a cylindrical lens above a plate with
lens parameters $L=4\,$mm, $R=12\,$mm. In Ref.~\cite{47}
this anomalous behavior of the electric force was
attributed to unavoidable deviations of the mechanically
polished and ground surfaces from the perfect spherical shape
assumed in elecrodynamical calculations. As recognized in
Ref.~\cite{46}, the same effect might be responsible for the
calibration problems arising in the case of a
centimeter-size cylinder above a plate. Because of this, the
consideration of much smaller cylinders seems to be preferable.

In this paper we investigate the possibility of combining the
advantage of high precision provided by the setup of a
micromachined oscillator and of a cylinder-plate configuration
with a cylinder radius of about $100\,\mu$m. For this purpose
we consider the thermal Casimir interaction between a metallized
plate and a microfabricated metal
coated cylinder attached to
a micromachined oscillator.
The potentially increased precision of force
measurements combined with the
 significantly decreased sizes of cylinders,
as compared with Refs.~\cite{41,46}, requires computations of the
thermal Casimir force with a more sophisticated account of such
factors as the lack of parallelism of the cylinder and plate and
the finiteness of the length of a cylinder. Hncre, we derive the
Lifshitz-type formulas for the Casimir force and for its
gradient for a cylinder and a plate made of real materials
using the proximity force approximation (PFA). The accuracy of
the results is determined by a comparison with the exact
expressions available for a cylinder and a plate made of
ideal metal \cite{43}.
Computations of the thermal correction to the Casimir force
and to its gradient using the resulting formulas are performed
in the framework of different theoretical approaches proposed
in the literature. The correction to the Casimir force and its
gradient due to the lack of parallelism of a plate and a cylinder is
determined in the nonmultiplicative way. The error due to the
finiteness of the length of a cylinder is estimated using the
results obtained with the help of world-line numerics \cite{48}.
Both static and dynamic experiments measuring the Casimir
force and its gradient in the cylinder-plate configuration
using a micromachined oscillator are proposed. It is shown
that the static experiment is better adapted for the
measurement of the thermal correction to the Casimir force.

The paper is organized as follows. In Sec.~II we derive the
Lifshitz-type formulas for a cylinder-plate configuration.
Section~III is devoted to the computations of thermal
corrections to the Casimir force and to its gradient.
The correction due to the lack of parallelism of a cylinder
and a plate is found in Sec.~IV. In Sec.~V we estimate
the error introduced by the finiteness of a cylinder.
The proposed experiments are discussed in Sec.~VI. Section~VII
contains our conclusions and discussion.

\section{Lifshitz-type formulas for a cylinder above a plate
made of real materials}

Let us consider the upper surface of a metallized plate
nearest to the cylinder as a coordinate plane
$z=0$.
For convenience of calculations, in Secs.~II--V we consider the
cylinder arranged above the plate. All the results obtained,
however, refer equally well to the cylinder arranged below the plate
(see Sec.~VI).
In real experiments, the cylinders used may have elliptical
cross-sections. Elliptical cylinders can
 be considered using the same methods,
 as applied below to circular cylinders.
The cylinder axis (coinciding with $y$ axis) is parallel
to the plane $z=0$ at a separation distance $R+a$, where $R$ is the
cylinder radius, and $a$ is the minimum separation between the
cylinder surface and the plate. The lower half of the cylinder
surface is described by the equation
\begin{equation}
z(x)=R+a-\sqrt{R^2-x^2}.
\label{eq1}
\end{equation}
\noindent
We approximately replace this half with a set of thick plane
plates extended along the $y$ axis
made of the same material as the cylinder and parallel
to the oscillator plate. Let us suppose that
the Casimir pressure $P\left(z(x),T\right)$
between each of these plates and its projection on the plane $z=0$
at temperature $T$ is  known.
According to the Derjaguin method \cite{6,49}, under the
condition $a\ll R$ the Casimir force acting between
a cylinder and a plate can be approximately calculated as
\begin{equation}
F(a,T)=\int_{\sigma}P\left(z(x),T\right)\,dx\,dy,
\label{eq2}
\end{equation}
\noindent
where integration holds over the projection of a cylinder on
the plane $z=0$. Incorporating some additional assumptions, the Derjaguin
method was reformulated as the PFA \cite{50} and widely used in
the literature for the calculation of the Casimir force between
a sphere and a plate (different versions of this method are
discussed in Refs.~\cite{6,51}).

We assume that the material of the plate and the cylinder (Au) is
characterized by the frequency-dependent dielectric permittivity
$\varepsilon(\omega)$. (In fact, as described in Sec.~VI, both
the plate and the cylinder are coated with an Au layer with
sufficient thickness such that they can be considered as made of
bulk Au.) The Casimir pressure between the plates at a separation
$z(x)$ at temperature $T$ is given by the Lifshitz formula
\begin{eqnarray}
&&
P\left(z(x),T\right)=-\frac{k_BT}{\pi}
\sum\limits_{l=0}^{\infty}{\vphantom{\sum}}^{\prime}
\int_{0}^{\infty}q_lk_{\bot}dk_{\bot}
\nonumber \\
&&~~~~~~~
\times
\sum\limits_{\alpha}
\frac{r_{\alpha}^2e^{-2q_lz(x)}}{1-r_{\alpha}^2e^{-2q_lz(x)}}.
\label{eq3}
\end{eqnarray}
\noindent
Here, $k_B$ is the Boltzmann constant, $k_{\bot}$ is the
projection of the wave vector on the plane $z=0$,
$q_l=(k_{\bot}^2+\xi_l^2/c^2)^{1/2}$, and
$\xi_l=2\pi k_B Tl/\hbar$ with $l=0,\,1,\,2,\,\ldots$ are the
Matsubara frequencies.
 The primed summation
means that the term with $l=0$ is multiplied by 1/2.
The reflection coefficients $r_{\alpha}$ for the
two polarizations of the electromagnetic field (transverse magnetic
with $\alpha={\rm TM}$ and transverse electric
with $\alpha={\rm TE}$) are given by
\begin{eqnarray}
&&
r_{\rm TM}=r_{\rm TM}(i\xi_l,k_{\bot})=
\frac{\varepsilon_lq_l-k_l}{\varepsilon_lq_l+k_l},
\nonumber \\
&&
r_{\rm TE}=r_{\rm TE}(i\xi_l,k_{\bot})=
\frac{q_l-k_l}{q_l+k_l},
\label{eq4}
\end{eqnarray}
\noindent
where
$k_l=\left[k_{\bot}^2+\varepsilon_l{\xi_l^2}/{c^2}\right]^{1/2}$
and $\varepsilon_l=\varepsilon(i\xi_l)$.
Notice that Eq.~(\ref{eq3}) can be identically represented as
\begin{eqnarray}
&&
P\left(z(x),T\right)=-\frac{k_BT}{\pi}
\sum\limits_{l=0}^{\infty}{\vphantom{\sum}}^{\prime}
\int_{0}^{\infty}q_lk_{\bot}dk_{\bot}
\nonumber \\
&&~~~~~~
\times
\sum\limits_{n=1}^{\infty}\sum\limits_{\alpha}
r_{\alpha}^{2n}e^{-2nq_lz(x)}.
\label{eq5}
\end{eqnarray}

Now we substitute Eq.~(\ref{eq5}) into Eq.~(\ref{eq2}) and arrive
at the expression
\begin{eqnarray}
&&
F(a,T)=-\frac{2k_BTL}{\pi}
\sum\limits_{l=0}^{\infty}{\vphantom{\sum}}^{\prime}
\int_{0}^{\infty}q_lk_{\bot}dk_{\bot}
\nonumber \\
&&~~~~\times
\sum\limits_{n=1}^{\infty}
(r_{\rm TM}^{2n}+r_{\rm TE}^{2n})
\int_{0}^{R}d\,xe^{-2nq_lz(x)},
\label{eq6}
\end{eqnarray}
\noindent
where $L$ is the length of the cylinder which is assumed
to be infinitely large. Using Eq.~(\ref{eq1}) we replace
the integration with respect to $x$ with an
integration with respect to $z$. We then introduce two
new integration variables, $u=(z-a)/R$ and $v=2q_la$, and
rewrite Eq.~(\ref{eq6}) in the form
\begin{eqnarray}
&&
F(a,T)=-\frac{k_BTRL}{4\pi a^3}
\sum\limits_{l=0}^{\infty}{\vphantom{\sum}}^{\prime}
\int_{\tau l}^{\infty}v^2\,dv
\label{eq7} \\
&&~~~~\times
\sum\limits_{n=1}^{\infty}
e^{-nv}(r_{\rm TM}^{2n}+r_{\rm TE}^{2n})
\int_{0}^{1}du\frac{(1-u)e^{-n\frac{R}{a}vu}}{\sqrt{1-(1-u)^2}}.
\nonumber
\end{eqnarray}
\noindent
Here, the dimensionless quantity $\tau$ is defined as
$\tau=4\pi k_B Ta/(\hbar c)$, and the reflection coefficients are
expressed in terms of dimensionless variables $v$ and
$\zeta_l=\xi_l/\omega_c=\tau l$, where $\omega_c=c/(2a)$,
in the following way:
\begin{eqnarray}
&&
r_{\rm TM}=r_{\rm TM}(v,\zeta_l)=
\frac{\varepsilon_lv-\sqrt{v^2+(\varepsilon_l-1)\zeta_l^2}}{\varepsilon_lv+
\sqrt{v^2+(\varepsilon_l-1)\zeta_l^2}},
\nonumber \\
&&
r_{\rm TE}=r_{\rm TE}(v,\zeta_l)=
\frac{v-\sqrt{v^2+(\varepsilon_l-1)\zeta_l^2}}{v+
\sqrt{v^2+(\varepsilon_l-1)\zeta_l^2}}.
\label{eq8}
\end{eqnarray}

Within the PFA, only the term of leading order in $a/R$ is
physically meaningful. Because of this the integral with respect
to $u$ in Eq.~(\ref{eq7}) can be evaluated as
\begin{eqnarray}
&&
\int_{0}^{1}du\frac{(1-u)}{\sqrt{1-(1-u)^2}}\,e^{-n\frac{R}{a}vu}
\nonumber \\
&&~~~~
=\int_{0}^{1}du\,e^{-n\frac{R}{a}vu}
\left[\frac{1}{\sqrt{2u}}-\frac{3}{4}\sqrt{\frac{u}{2}}+
O(u^{3/2})\right]
\nonumber \\
&&~~~~
=\sqrt{\frac{\pi a}{2nvR}}+O\left[\left(\frac{a}{R}\right)^{3/2}\right].
\label{eq9}
\end{eqnarray}
\noindent
Substituting Eq.~(\ref{eq9}) into Eq.~(\ref{eq7}) we obtain
\begin{eqnarray}
&&
F(a,T)=-\frac{k_BTL}{4\sqrt{\pi} a^2}\sqrt{\frac{R}{2a}}
\sum\limits_{l=0}^{\infty}{\vphantom{\sum}}^{\prime}
\int_{\tau l}^{\infty}v^{3/2}\,dv
\nonumber \\
&&~~~~\times
\sum\limits_{n=1}^{\infty}
\frac{e^{-nv}}{\sqrt{n}}(r_{\rm TM}^{2n}+r_{\rm TE}^{2n}).
\label{eq10}
\end{eqnarray}
\noindent
Using the definition of the polylogarithm function, the
Lifshitz-type formula (\ref{eq10}) describing the Casimir force
for a cylinder in close proximity to a plate can be
represented in the form
\begin{eqnarray}
&&
F(a,T)=-\frac{k_BTL}{4\sqrt{\pi} a^2}\sqrt{\frac{R}{2a}}
\sum\limits_{l=0}^{\infty}{\vphantom{\sum}}^{\prime}
\int_{\tau l}^{\infty}v^{3/2}\,dv
\nonumber \\
&&~~~~\times
\left[{\rm Li}_{1/2}(r_{\rm TM}^{2}e^{-v})+
{\rm Li}_{1/2}(r_{\rm TE}^{2}e^{-v})\right].
\label{eq11}
\end{eqnarray}
\noindent
This equation can be used in numerical computations of the
Casimir force acting between a cylinder and a plate (see the
next section).

In the case of a cylinder and a plate made of ideal metal,
$r_{\rm TM}^2=r_{\rm TE}^2=1$. Then at zero temperature
Eq.~(\ref{eq11}) takes the form
\begin{equation}
F^{\rm IM}(a,0)=-\frac{k_BTL}{2\sqrt{\pi} a^2}\sqrt{\frac{R}{2a}}
\int_{0}^{\infty}dl
\int_{\tau l}^{\infty}v^{3/2}\,dv
\sum\limits_{n=1}^{\infty}
\frac{e^{-nv}}{\sqrt{n}}.
\label{eq12}
\end{equation}
\noindent
Introducing the new variable $\zeta=\tau l$, changing the order of
integrations and calculating the integrals with respect to $\zeta$
and $v$ one obtains
\begin{equation}
F^{\rm IM}(a,0)=-\frac{15\hbar cL}{64\pi a^3}\sqrt{\frac{R}{2a}}
\sum\limits_{n=1}^{\infty}
\frac{1}{n^4}=
-\frac{\pi^3\hbar cL}{384 a^3}\sqrt{\frac{R}{2a}}.
\label{eq13}
\end{equation}
\noindent
Equation (\ref{eq13}) coincides with the familiar expression
for the configuration of an ideal metal
cylinder above an ideal metal plate obtained
using the PFA in Ref.~\cite{41A} (see also Refs.~\cite{6,41}).

Both Eqs.~(\ref{eq11}) and (\ref{eq13}) are approximate results.
Bearing in mind that Eq.~(\ref{eq11}) can be used for the comparison
of experimental data with theory, it is important to estimate
how accurate it is. Fortunately, for an ideal metal cylinder
above an ideal metal plate at zero temperature the exact
analytical result
for the two leading terms of the expansion of the Casimir force
in powers of $a/R$ is available \cite{43}
\begin{eqnarray}
&&
F_{\rm ex}^{\rm IM}(a,0)=
-\frac{\pi^3\hbar cL}{384\pi a^3}\sqrt{\frac{R}{2a}}
\left(1-C_{\rm ex}\frac{a}{R}\right),
\nonumber \\
&&
C_{\rm ex}=\frac{4}{\pi^2}-\frac{7}{60}\approx 0.2886.
\label{eq14}
\end{eqnarray}
\noindent
This result was confirmed by means of numerical computations in
Ref.~\cite{44A}.
{}From the comparison of Eqs.~(\ref{eq13}) and (\ref{eq14}) it can
be seen that under the condition $a\ll R$ the deviation of the PFA
result from the exact result does not exceed $0.3a/R$.
For a typical experimental value of $a/R\approx 10^{-3}$
the error in using the PFA turns out to be of about 0.03\%.
For the configuration of
an ideal metal sphere above an ideal metal plate at temperature
$k_BT\gg\hbar c/R$ it was shown \cite{52} that in the zeroth order
of $a/R$ the PFA leads to exact results for the zero-temperature
contribution to the Casimir force and, separately, for the
thermal correction to it. It was also demonstrated that at
zero temperature for a sphere and a plate made of real metals
the deviation of the PFA results from the exact results is
smaller than for a sphere and a plate made of an ideal metal
\cite{53,54,55,56}. Finally, exact computations of the thermal
Casimir force at nonzero temperature in a sphere-plate
configuration performed with different $a/R$ show that the exact
results approach the PFA results when $a/R\to 0$
\cite{56,57,58}. All the above suggests that for the configuration
of a real metal cylinder above a real metal plate, the error of the
Lifshitz-type Eq.~(\ref{eq11}) obtained using the PFA is of about
the same, $0.3a/R$, as holds for ideal metal bodies at $T=0$.

The Lifshitz-type formula for the gradient of the Casimir force
can be obtained in a similar way to Eq.~(\ref{eq11}).
For this purpose we differentiate Eq.~(\ref{eq6}) with respect
to $a$ taking into account Eq.~(\ref{eq1}),
\begin{eqnarray}
&&
F^{\prime}(a,T)\equiv\frac{\partial F(a,T)}{\partial a}
=\frac{4k_BTL}{\pi}
\sum\limits_{l=0}^{\infty}{\vphantom{\sum}}^{\prime}
\int_{0}^{\infty}q_l^2k_{\bot}dk_{\bot}
\nonumber \\
&&~~~~\times
\sum\limits_{n=1}^{\infty}n
(r_{\rm TM}^{2n}+r_{\rm TE}^{2n})
\int_{0}^{R}d\,xe^{-2nq_lz(x)}.
\label{eq15}
\end{eqnarray}
\noindent
Dealing with this equation as with Eq.~(\ref{eq6}), in the leading
order of $a/R$ one obtains
\begin{eqnarray}
&&
\frac{\partial F(a,T)}{\partial a}=
\frac{k_BTL}{4\sqrt{\pi} a^3}\sqrt{\frac{R}{2a}}
\sum\limits_{l=0}^{\infty}{\vphantom{\sum}}^{\prime}
\int_{\tau l}^{\infty}v^{5/2}\,dv
\nonumber \\
&&~~~~\times
\sum\limits_{n=1}^{\infty}
\sqrt{n}e^{-nv}(r_{\rm TM}^{2n}+r_{\rm TE}^{2n}).
\label{eq16}
\end{eqnarray}
\noindent
Using the definition of the polylogarithm function
results in
\begin{eqnarray}
&&
\frac{\partial F(a,T)}{\partial a}=
\frac{k_BTL}{4\sqrt{\pi} a^3}\sqrt{\frac{R}{2a}}
\sum\limits_{l=0}^{\infty}{\vphantom{\sum}}^{\prime}
\int_{\tau l}^{\infty}v^{5/2}\,dv
\nonumber \\
&&~~~~\times
\left[{\rm Li}_{-1/2}(r_{\rm TM}^{2}e^{-v})+
{\rm Li}_{-1/2}(r_{\rm TE}^{2}e^{-v})\right].
\label{eq17}
\end{eqnarray}
\noindent
For an ideal metal cylinder above an ideal metal plate at
zero temperature Eq.~(\ref{eq17}) leads to the result
\begin{equation}
\frac{\partial F^{\rm IM}(a,0)}{\partial a}=
\frac{7\pi^3\hbar cL}{768 a^4}\sqrt{\frac{R}{2a}},
\label{eq18}
\end{equation}
\noindent
which is also immediately obtainable by the differentiation
of Eq.~(\ref{eq13}) with respect to $a$.

In the high-temperature (large-separation) limit only the
zero-frequency terms of Eqs.~(\ref{eq11}) and (\ref{eq17})
contribute to the result. Specifically, for the ideal metal case
one obtains from Eq.~(\ref{eq11})
\begin{eqnarray}
F^{\rm IM}(a,T)&=&-\frac{k_BTL}{4\sqrt{\pi} a^2}\sqrt{\frac{R}{2a}}
\int_{0}^{\infty}v^{3/2}
{\rm Li}_{1/2}(e^{-v})\,dv
\nonumber \\
&=&-\frac{3\zeta(3)k_BTL}{16 a^2}\sqrt{\frac{R}{2a}},
\label{eq19}
\end{eqnarray}
\noindent
where $\zeta(z)$ is the Riemann zeta function.
For the gradient of the Casimir force this results in
\begin{equation}
\frac{\partial F^{\rm IM}(a,T)}{\partial a}=
\frac{15\zeta(3)k_BTL}{32 a^3}\sqrt{\frac{R}{2a}}.
\label{eq20}
\end{equation}
\noindent
For real metals the asymptotic behavior at high temperature
depends on the assumed model of dielectric properties
(see the next section).

Although the immediately planned experiments are for metallic
test bodies, the case of a dielectric cylinder above a
dielectric plate is also of some experimental interest.
For dielectrics with a finite dielectric permittivity at zero
frequency, $\varepsilon(0)=\varepsilon_0$, the high-temperature
behavior of the Casimir force is simply obtainable from
Eqs.~(\ref{eq10}) and (\ref{eq11}).
In this case at zero frequency we have
\begin{equation}
r_{\rm TM}(0,v)=\frac{\varepsilon_0-1}{\varepsilon_0+1}=r_0,
\quad
r_{\rm TE}(0,v)=0.
\label{eq21}
\end{equation}
\noindent
The zero-frequency term of Eq.~(\ref{eq10}) then takes the
form
\begin{equation}
F^{\rm diel}(a,T)=-\frac{k_BTL}{8\sqrt{\pi}a^2}
\sqrt{\frac{R}{2a}}\sum\limits_{n=1}^{\infty}
\frac{r_0^{2n}}{\sqrt{n}}
\int_{0}^{\infty}v^{3/2}e^{-nv}dv.
\label{eq22}
\end{equation}
\noindent
Calculating the integral with respect to $v$ the high-temperature
behavior of the Casimir force between a dielectric plate and
dielectric cylinder is obtained:
\begin{eqnarray}
F^{\rm diel}(a,T)&=&-\frac{3k_BTL}{32 a^2}\sqrt{\frac{R}{2a}}
\sum\limits_{n=1}^{\infty}\frac{r_0^{2n}}{n^3}
\nonumber \\
&=&-\frac{3k_BTL}{32 a^2}\sqrt{\frac{R}{2a}}
{\rm Li}_3(r_0^2).
\label{eq23}
\end{eqnarray}
\noindent
The same equation, where $r_0^2$ is replaced with $r_0$, describes
the high-temperature Casimir force acting between a dielectric
plate and metallic cylinder (or, equivalently, metallic plate and
dielectric cylinder). The latter result does not depend on the
model used for the description of dielectric properties of the metal.

\section{Computations of the thermal correction to the Casimir
force between a cylinder and a plate coated with gold}

Here, we use Eqs.~(\ref{eq11}) and (\ref{eq17}) for computations
of the relative thermal corrections to the Casimir force and to
the gradient of the Casimir force defined as
\begin{eqnarray}
&&
\delta_{T}^{(1)}=\frac{F(a,T=300\,K)-F(a,0)}{F(a,T=300\,K)},
\label{eq24} \\
&&
\delta_{T}^{(2)}=\frac{F^{\prime}(a,T=300\,K)-
F^{\prime}(a,0)}{F^{\prime}(a,T=300\,K)}.
\nonumber
\end{eqnarray}
\noindent
This requires knowledge of the complete data for the dielectric
permittivity of Au along the imaginary frequency axis.
Usually these are found using the dispersion relation
\begin{equation}
\varepsilon(i\xi)=1+\frac{2}{\pi}\int_{0}^{\infty}
\frac{\omega\,{\rm Im}\,\varepsilon(\omega)}{\omega^2+\xi^2}
d\omega,
\label{eq25}
\end{equation}
\noindent
where ${\rm Im}\,\varepsilon(\omega)$ is calculated from real
and imaginary parts of the complex index of refraction $n$,
tabulated, e.g., in Ref.~\cite{59}.
Keeping in mind that ${\rm Re}\,n(\omega)$ and ${\rm Im}\,n(\omega)$
are known only within a restricted frequency range (from 0.125\,eV
to 10000\,eV for Au in Ref.~\cite{59}) the problem
of extrapolation of ${\rm Im}\,\varepsilon(\omega)$ to lower and
higher frequencies arises (in fact only the extrapolation to lower
frequencies is of practical importance).
According to the Drude model approach \cite{6,11,17,17a,18,19},
the extrapolation of ${\rm Im}\,\varepsilon(\omega)$ to lower
frequencies is performed using the imaginary part of the
dielectric permittivity
\begin{equation}
\varepsilon_{D}(\omega)=1-\frac{\omega_p^2}{\omega(\omega+i\gamma)},
\label{eq26}
\end{equation}
\noindent
where $\omega_p$ is the plasma frequency and $\gamma$ is the
relaxation parameter (for Au we use \cite{60}
$\omega_p=9.0\,$eV and $\gamma=0.035\,$eV).

According to the plasma model approach \cite{6,11,34,61,62},
the dielectric permittivity can be expressed in the following
oscillator form
\begin{equation}
\varepsilon(\omega)=1-\frac{\omega_p^2}{\omega^2}
+\sum\limits_{j=1}^{N}
\frac{g_j}{\omega_j^2-\omega^2-i\gamma_j\omega},
\label{eq27}
\end{equation}
\noindent
where the parameters of oscillators $\omega_j\neq 0,\>\gamma_j$,
and $g_j$ are determined \cite{34} from the best fit of
${\rm Im}\,\varepsilon(\omega)$ to the tabulated optical data
for $n$ related to interband transitions of core electrons.
As discussed in Sec.~I, the Drude and the plasma model approaches
lead to drastically different predictions for the thermal
correction to the Casimir force in the sphere-plate geometry.
Here, we perform all computations using both approaches in order
to find additional evidence which
could help to resolve this difference. Note that
simple Drude and plasma models, which do not take interband
transitions into account, were used in Ref.~\cite{41A} to compute
the Casimir interaction in a cylinder-plate configuration at
nonzero temperature.
Recently an
alternative approach was also proposed \cite{63}. It aimed
to determine $\varepsilon(i\xi)$ from both
${\rm Im}\,\varepsilon(\omega)$ and ${\rm Re}\,\varepsilon(\omega)$
within a frequency range where they are measured without using
any extrapolations, but assuming the Drude-type behavior at zero
frequency. It was shown \cite{64}, however, that large errors in the
measurement data for ${\rm Re}\,\varepsilon(\omega)$ complicate
the application of this approach in practical computations.

In Fig.~1 we present the computational results for the quantities
$\delta_{T,D}^{(1)}$ (line 1) and
$\delta_{T,D}^{(2)}$ (line 2) as functions of separation
in the region from 0.1 to $5\,\mu$m. The computations were performed
using the Drude model approach. As can be seen in Fig.~1, the
thermal corrections to both the Casimir force and its gradient in
a cylinder-plate configuration are negative over a wide range of
separations. This is typical for a Drude model approach in the
configurations of two parallel plates and a sphere above a plate
as well. It should also be  noted that the relative thermal correction
achieves rather large values at short separation distances.
Specifically, at experimentally relevant separations
$a=150$, 200, 300, 500, and 750\,nm the thermal correction to
the Casimir force is equal to --1.8\%, --2.7\%, --4.6\%, --8.6\%,
and --13.9\%, respectively. At the same respective separations the
thermal correction to the gradient of the Casimir force achieves
more moderate values --0.1\%, --1.6\%, --2.9\%, --5.6\%, and --9.3\%.
The magnitude of the relative thermal correction to
the Casimir force $|\delta_{T,D}^{(1)}|$ achieves its maximum
value 41.6\% at $a=2.55\,\mu$m, while the maximum magnitude of the
thermal correction to the gradient of the Casimir force
$|\delta_{T,D}^{(2)}|=52$\% occurs at $a=3.6\mu$m. Thus, at large
separation distances, dynamic experiments, where $\partial F/\partial a$
is a directly measured quantity, are most suitable for the
detection of the thermal correction predicted within the Drude
model approach.

In the high-temperature (large-separation) limit it is easy to
obtain the analytic expression for the thermal Casimir force
calculated using the Drude model approach. Taking into account
that Eqs.~(\ref{eq8}) and (\ref{eq26}) lead to
$r_{\rm TM}(0,v)=1$ and $r_{\rm TE}(0,v)=0$, the contribution
of the zero-frequency term in Eq.~(\ref{eq11}) is given by
\begin{eqnarray}
F^{D}(a,T)&=&-\frac{k_BTL}{8\sqrt{\pi} a^2}\sqrt{\frac{R}{2a}}
\int_{0}^{\infty}v^{3/2}
{\rm Li}_{1/2}(e^{-v})\,dv
\nonumber \\
&=&-\frac{3\zeta(3)k_BTL}{32 a^2}\sqrt{\frac{R}{2a}}.
\label{eq28}
\end{eqnarray}
\noindent
For the gradient of the Casimir force
\begin{equation}
\frac{\partial F^{\rm D}(a,T)}{\partial a}=
\frac{15\zeta(3)k_BTL}{64 a^3}\sqrt{\frac{R}{2a}}
\label{eq29}
\end{equation}
\noindent
is obtained.

We next present the computational results in the case when the
plasma model approach is applied to describe metallic coatings
on a cylinder and a plate. In Fig.~2(a) the computational results
for the quantities
$\delta_{T,p}^{(1)}$ (line 1) and
$\delta_{T,p}^{(2)}$ (line 2) are shown as functions of
separation in the region from 0.1 to $5\,\mu$m.
In Fig.~2(b) the same results in the region from 0.1 to $1\,\mu$m
are shown at a larger scale. As for the configurations
of two parallel plates and a sphere above a plate, the thermal
correction for a cylinder above a plate is positive when the
plasma model approach is used. As can be seen in Fig.~2(a,b),
at short separations the relative thermal correction obtained
within the plasma model approach is negligibly small. Thus,
the correction to the Casimir force is equal to 0.016\%, 0.024\%,
0.044\%, 0.13\%, and 0.38\% at separations $a=150$, 200, 300, 500,
and 750\,nm, respectively. At the same respective separations,
the thermal correction to the gradient of the Casimir force is
equal to 0.0014\%, 0.0023\%, 0.0046\%, 0.012\%, and 0.029\%,
i.e., it is even less. This makes the observation of the thermal
correction at short separations, as predicted by the plasma model,
impossible in either the static or
dynamic experiments. However, at larger separations the
thermal correction to the Casimir force computed using the
plasma model approach quickly increases. It is equal to 0.9\%,
7.2\%, and 46\% at separations $a=1$, 2, and $5\,\mu$m,
respectively. At the same respective separations,
the thermal correction to the gradient of the Casimir force is
equal to 0.063\%, 0.82\%, and 26.7\%. Thus, within the plasma
model approach the relative thermal correction to the Casimir
force is always larger than to the gradient of the Casimir
force.

As in the case of the Drude model approach, one can obtain
analytic  expressions for the thermal Casimir force and its
gradient at high temperature. {}From Eqs.~(\ref{eq8}) and
(\ref{eq27}), for the reflection coefficients at zero frequency,
one obtains
\begin{eqnarray}
&&
r_{\rm TM}(0,v)=1,
\label{eq30} \\
&&
r_{\rm TE}(0,v)=
\frac{\alpha v-\sqrt{\alpha^2v^2+1}}{\alpha v+\sqrt{\alpha^2v^2+1}}
=-(\alpha v-\sqrt{\alpha^2v^2+1})^2,
\nonumber
\end{eqnarray}
\noindent
where $\alpha=\omega_c/\omega_p\equiv\delta_0/(2a)$. The quantity
$\delta_0=c/\omega_p$ has the meaning of the skin depth in the
frequency region of infrared optics. Equation~(\ref{eq11}) then
leads to
\begin{eqnarray}
&&
F^{p}(a,T)=-\frac{k_BTL}{8\sqrt{\pi} a^2}\sqrt{\frac{R}{2a}}
\int_{0}^{\infty}v^{3/2}\,dv
\label{eq31} \\
&&~~~~~~~~~
\times
\left[{\rm Li}_{1/2}\left(e^{-v}\right)+
{\rm Li}_{1/2}\left(r_{\rm TE}^2(0,v)e^{-v}\right)\right].
\nonumber
\end{eqnarray}
\noindent
The integral of the first term on the right-hand side of
of Eq.~(\ref{eq31}) was calculated in Eq.~(\ref{eq28}).
The integral of the second term can be calculated
approximately with the help of perturbation theory in powers of
the small parameter $\alpha$ (for Au
$\delta_0\approx 22\,\mbox{nm}\ll a$). Using Eq.~(\ref{eq30}), one obtains
\begin{equation}
r_{\rm TE}^{2n}(0,v)=1-4nv\alpha+8n^2v^2\alpha^2+O(\alpha^3).
\label{eq32}
\end{equation}
\noindent
The second integral on the right-hand side of
of Eq.~(\ref{eq31}) is then found using this equation:
\begin{eqnarray}
&&
\int_{0}^{\infty}v^{3/2}
{\rm Li}_{1/2}\left(r_{\rm TE}^2(0,v)e^{-v}\right)dv
\nonumber \\
&&~~~~~
=\sum\limits_{n=1}^{\infty}\frac{1}{\sqrt{n}}
\int_{0}^{\infty}v^{3/2}(1-4n\alpha v+8n^2\alpha^2 v^2)
e^{-nv}dv
\nonumber \\
&&~~~~~
=\frac{3\sqrt{\pi}\zeta(3)}{4}(1-10\alpha+70\alpha^2)
\label{eq33} \\
&&~~~~~
=\frac{3\sqrt{\pi}\zeta(3)}{4}\left[1-
5\frac{\delta_0}{a}+\frac{35}{2}\left(\frac{\delta_0}{a}
\right)^{\!2}\right].
\nonumber
\end{eqnarray}
\noindent
By adding the contribution of the first integral
in Eq.~(\ref{eq31}), we obtain the high-temperature asymptotic
behavior of the Casimir force
\begin{equation}
F^{p}(a,T)=-\frac{3\zeta(3)k_BTL}{16a^2}\sqrt{\frac{R}{2a}}
\left[1-
\frac{5}{2}\,\frac{\delta_0}{a}+\frac{35}{4}
\left(\frac{\delta_0}{a}\right)^{\!2}\right].
\label{eq34}
\end{equation}
\noindent
The high-temperature asymptotic behavior for the gradient of the
Casimir force is given by
\begin{equation}
\frac{\partial F^{p}(a,T)}{\partial a}=
\frac{15\zeta(3)k_BTL}{32a^3}\sqrt{\frac{R}{2a}}
\left[1-
\frac{7}{2}\,\frac{\delta_0}{a}+\frac{63}{4}
\left(\frac{\delta_0}{a}\right)^{\!2}\right].
\label{eq35}
\end{equation}

{}From the comparison of Eqs.~(\ref{eq34}) and (\ref{eq35}) with
Eqs.~(\ref{eq28}) and (\ref{eq29}), respectively, it can be seen
that under the condition $a\ll R$ the main contributions to the
high-temperature Casimir force and its gradient calculated using
the plasma model approach are two times larger than the
Casimir force and its gradient calculated using
the Drude model approach.
The same holds in the configuration of two parallel plates and
for a sphere above a plate under the condition $a\ll R$
\cite{6,11}.

\section{Correction due to lack of parallelism of a plate and a
cylinder}

Experimentally it is impossible to preserve the exact parallelism
of a plate and a cylinder as was always assumed in the above
calculations. Because of this, it is important to find the
correction to the derived results which arises if there is
some small angle $\theta\ll 1$ between the plate and the axis
of a cylinder.  For ideal metal bodies such a correction factor
which depends on the mean separation between the cylinder and
plate
was found in Ref.~\cite{41}. The magnitude of the Casimir force
between real metal cylinder and a plate
taking into account their
nonparallelity
was also estimated by multiplying the result obtained for
parallel bodies by the same factor.
Calculations show, however, that for real metal bodies
corrections due to deviations from perfect geometry (for
instance, due to surface roughness or due to a nonparallelity
of two surfaces) are not of precisely multiplicative
character \cite{6,11}. Specifically, at short separations
(typically at $a<200\,$nm) the multiplicative approach may
lead to large errors. Because of this, we here calculate
the thermal Casimir force between a cylinder and a plate
in a more inclusive way by applying the PFA to account
not only for a cylindrical geometry but for a nonparallelity
of a cylinder and a plate as well.

We start with Eq.~(\ref{eq6}) for the Casimir force between
a plate and a cylinder parallel to it.
It is necessary to take into account
that the cylinder is now inclined by a small angle $\theta$
with respect to the plane $z=0$.
This means that thick plane plates extended along the $y$ axis,
which are used in Sec.~II as a substitution for a cylinder in
the PFA, are also inclined at an angle $\theta$.
We replace each of these inclined plates
with a set of plates of area $dx\,dy$ parallel to
the plane $z=0$. This is equivalent to replacing
$z(x)$ in Eq.~(\ref{eq6}) with $z(x)+\theta y$, and
subsequently
averaging over the length of the cylinder. The Casimir
force then takes the form
\begin{eqnarray}
&&
F(a,T,\theta)=-\frac{2k_BT}{\pi}
\sum\limits_{l=0}^{\infty}{\vphantom{\sum}}^{\prime}
\int_{0}^{\infty}q_lk_{\bot}dk_{\bot}
\label{eq36} \\
&&~~\times
\sum\limits_{n=1}^{\infty}
(r_{\rm TM}^{2n}+r_{\rm TE}^{2n})
\int_{0}^{R}d\,x\int_{-L/2}^{L/2}dy
e^{-2nq_l[z(x)+\theta y]},
\nonumber
\end{eqnarray}
\noindent
where $a$ is now the mean minimum separation between the cylinder
and the plate. At $\theta=0$ Eq.~(\ref{eq36}) coincides with
Eq.~(\ref{eq6}).

The integration with respect to $y$ in Eq.~(\ref{eq36}) is
trivial. We then introduce the other integration variables
$u$ and $v$,
defined in Sec.~II, and perform calculations similar to those
leading to Eq.~(\ref{eq10}). Finally, retaining only the leading
order contribution in the small parameter $a/R$, we arrive at
\begin{eqnarray}
&&
F(a,T,\theta)=-\frac{k_BTL}{4\sqrt{\pi} a^2}\sqrt{\frac{R}{2a}}
\sum\limits_{l=0}^{\infty}{\vphantom{\sum}}^{\prime}
\int_{\tau l}^{\infty}v^{3/2}\,dv
\label{eq37} \\
&&~~~~\times
\sum\limits_{n=1}^{\infty}
\frac{e^{-nv}}{\sqrt{n}}(r_{\rm TM}^{2n}+r_{\rm TE}^{2n})
\frac{\sinh(A_{\theta}nv)}{A_{\theta}nv},
\nonumber
\end{eqnarray}
\noindent
where one additional small parameter is defined as $A_{\theta}=\theta
L/(2a)$. For ideal metals at zero temperature Eq.~(\ref{eq36})
reads as
\begin{eqnarray}
&&
F^{\rm IM}(a,0,\theta)=-\frac{\hbar cL}{8\pi\sqrt{\pi} a^3}\sqrt{\frac{R}{2a}}
\int_{0}^{\infty}v^{5/2}\,dv
\nonumber \\
&&~~~~~~~\times
\sum\limits_{n=1}^{\infty}
\frac{e^{-nv}}{\sqrt{n}}
\frac{\sinh(A_{\theta}nv)}{A_{\theta}nv}.
\label{eq38}
\end{eqnarray}
\noindent
After performing the integration and summation, Eq.~(\ref{eq38})
results in
\begin{equation}
F^{\rm IM}(a,0,\theta)=\kappa(A_{\theta})F^{\rm IM}(a,0),
\label{eq39}
\end{equation}
\noindent
where $F^{\rm IM}(a,0)$ holds for an ideal metal cylinder parallel to
an ideal metal plate, and is defined in Eq.~(\ref{eq13}), and the
correction factor due to nonparallelity is given by
\begin{equation}
\kappa(A_{\theta})=\frac{1}{5A_{\theta}}\left[
\frac{1}{(1-A_{\theta})^{5/2}}-
\frac{1}{(1+A_{\theta})^{5/2}}\right].
\label{eq40}
\end{equation}
\noindent
Note that this correction factor was first obtained
in Ref.~\cite{41} by direct application of the PFA to an
ideal metal cylinder inclined at an angle $\theta$ to an ideal
metal plate.
Here we have shown that Eq.~(\ref{eq40}) follows from  a more
general Lifshitz-type formula (\ref{eq38}) which takes into
account the nonmultiplicative effects due to nonparallelity
of a plate and a cylinder. The correction factor
$\kappa(A_{\theta})$ can be also used at nonzero temperature in order
to estimate the Casimir force between a cylinder and a plate
under an angle $\theta$ by means of the approximate
multiplicative approach
\begin{equation}
F^{\rm m}(a,T,\theta)=\kappa(A_{\theta})F(a,T).
\label{eq41}
\end{equation}
\noindent
Here we estimate the accuracy of this approach at different
angles and separation distances.

The same procedure, as described above for the Casimir force,
can be used in order to find a nonmultiplicative
effect of a nonzero angle between a cylinder and a plate
on the gradient of the Casimir force.
As a result, Eq.~(\ref{eq16}) is replaced by
\begin{eqnarray}
&&
\frac{\partial F(a,T)}{\partial a}=
\frac{k_BTL}{4\sqrt{\pi} a^3}\sqrt{\frac{R}{2a}}
\sum\limits_{l=0}^{\infty}{\vphantom{\sum}}^{\prime}
\int_{\tau l}^{\infty}v^{5/2}\,dv
\nonumber \\
&&~~~~\times
\sum\limits_{n=1}^{\infty}
\sqrt{n}e^{-nv}(r_{\rm TM}^{2n}+r_{\rm TE}^{2n})
\frac{\sinh(A_{\theta}nv)}{A_{\theta}nv}.
\label{eq42}
\end{eqnarray}

Using Eqs.~(\ref{eq11}) and (\ref{eq37}) we performed numerical
computations of the quantity
\begin{equation}
\kappa^{(\rm nm)}(a,A_{\theta},T)=
\frac{F(a,T,\theta)}{F(a,T)}
\label{eq43}
\end{equation}
\noindent
at different $a$ and $A_{\theta}$. This quantity takes into
account the nonmultiplicative effects due to a nonzero angle between
a cylinder and a plate.
When compared with $\kappa(A_{\theta})$, it allows us to
determine the application region of the multiplicative
approach. Note that $\kappa^{(\rm nm)}$ depends on separation
through the parameter $A_{\theta}$, and through the $a$-dependent
quantity $\tau$ in the lower integration limits
in Eqs.~(\ref{eq11}) and (\ref{eq37}). As for the correction
factor $\kappa$ defined in Eq.~(\ref{eq40}), it depends on
separation only through the parameter $A_{\theta}$.

The computational results for $\kappa^{(\rm nm)}$ (lines 1--6)
and $\kappa$ (line 7) at $T=300\,$K are presented in Table~I.
Column 1 indicates separations $a$ at which the values
of $\kappa^{(\rm nm)}$ are computed (from 100 to 500\,nm).
Columns 2,\,\,3,\,\,4, and 5 contain the results obtained
for $A_{\theta}=0.01$, 0.05, 0.1, and 0.5, respectively.
Keeping in mind the definition of $A_{\theta}$,
the value of $\theta=2aA_{\theta}/L$ in each column changes
with the change of separation. To determine the application
region of the multiplicative approach, we assume that this
approach should not lead to an error larger than the error due
to the PFA. For the parameters of a cylinder we take
$L=R=100\,\mu$m. Then, for instance, at $a=100\,$nm
the relative error of the PFA is equal to $0.3a/R=0.0003$
(see Sec.~II). If $A_{\theta}=0.01$ (see the column 2 of
Table~I) we have
$\kappa-\kappa^{(\rm nm)}(100\,\mbox{nm})=0.00006$, i.e., much
less than the error of the PFA. Keeping in mind that at
$a=100\,$nm the inequality $A_{\theta}\leq 0.01$ is certainly
valid at angles $\theta\leq 4^{\prime\prime}$, one arrives at the
conclusion that with these angles between a cylinder and a plate
the  multiplicative approach is applicable.
It is very hard, however, to ensure angles
$\theta\leq 4^{\prime\prime}\approx 20\,\mu$rad experimentally.
Because of this, at short separations the nonmultiplicative
approach should be used for the comparison between the
experimental data and theory.
With $A_{\theta}=0.01$ the multiplicative approach
is applicable at larger separations, as well as at even larger
angles. As one more example we consider $A_{\theta}=0.05$
(the column 3 of Table~I). Here,
$\kappa-\kappa^{(\rm nm)}(100\,\mbox{nm})=0.0015$, i.e., larger
than the error of the PFA at $a=100\,$nm. Thus, the
 multiplicative approach is not applicable. However, at
 $a=500\,$nm the error of the PFA is equal to 0.0015, whereas
$\kappa-\kappa^{(\rm nm)}(500\,\mbox{nm})=0.0006$. Thus, at
this separation with chosen parameters of a cylinder
the multiplicative approach is applicable
at $\theta\leq 1.7^{\prime}$. At larger angles the
nonmultiplicative results in Eqs.~(\ref{eq37}) and (\ref{eq42}) should
be used to compute the Casimir force and its gradient with
sufficient precision.

\section{Error due to finite length of a cylinder}

The Lifshitz-type formulas (\ref{eq11}) and (\ref{eq17}) were
obtained under an assumption that the length $L$ of the cylinder
is infinitely large. In fact these formulas define the Casimir
force and its gradient per unit length of a cylinder.
Keeping in mind that in our experimental setup
the length of the cylinder $L$ is not only finite
 but might be of the same order as the cylinder radius $R$, it is important
 to estimate the possible error due to the finiteness of $L$.
Here, we solve this problem for an ideal metal cylinder above an
ideal metal plate. The derived results can serve as a reasonable
estimation for real metal bodies as well.

We start with the case when the plate
 is sufficiently large that the projection of the
cylinder on the plane $(x,y)$ is situated
within its interior [see Fig.~3(a)].
This is the case for the proposed experiment (see Sec.~VI).
It is assumed that the length of plane
plates with an area $Ldx$, which replace a cylinder when we apply
the PFA to calculate the Casimir force (see Sec.~II), is finite.
As a result, two real edges of length $dx$ each are formed.
Because of this, it is not legitimate to use either the
original Casimir expression for the force per unit area
$-\pi^2\hbar c/(240a^4)$, or its Lifshitz counterpart (\ref{eq3})
applicable to real metals, in order to perform calculations in
the framework of the PFA. Instead, we apply the expression
\cite{48} for the Casimir force acting between an infinite plate
and a  finite plate parallel to it of area $S$ with edge length
$l$ spaced $z$ apart, obtained using the method of world line
numerics
\begin{equation}
F_{\rm ed}(z,0)=-\frac{\pi^2\hbar cS}{480z^4}-\gamma_a
\frac{\hbar cl}{z^3}.
\label{eq44}
\end{equation}
\noindent
This equation with $\gamma_a=5.23\times 10^{-3}$ was obtained
for a scalar field with Dirichlet boundary conditions on the
plates at zero temperature. Keeping in mind applications to the
electromagnetic field, the Casimir force acting between the
partial plate of area $Ldx$, replacing the cylindrical
surface in the PFA, and the plane plate
is given by
\begin{equation}
dF_{\rm ed}^{\rm IM}(z,0)=-\frac{\pi^2\hbar cLdx}{240z^4}-4\gamma_a
\frac{\hbar cdx}{z^3},
\label{eq45}
\end{equation}
\noindent
where $z=z(x)$ is defined in Eq.~(\ref{eq1}), and the length of
real edges is $l=2dx$.

The integration of Eq.~(\ref{eq45}) over the lower half of a
cylinder in the leading order of the small parameter $a/R$
results in
\begin{eqnarray}
&&
F_{\rm ed}^{\rm IM}(a,0)=\int_{-R}^{R}dF_{\rm ed}^{\rm IM}(z,0)
\label{eq46} \\
&&~~~~~
=-\frac{\pi^3\hbar cL}{384a^3}
\sqrt{\frac{R}{2a}} -
\frac{3\pi\gamma_a\hbar c}{a^2}\sqrt{\frac{R}{2a}}
\nonumber \\
&&~~~~~
=
F^{\rm IM}(a,0)\left(1+C_{\rm ed}\frac{a}{L}\right),
\nonumber
\end{eqnarray}
\noindent
where $F^{\rm IM}(a,0)$ is defined in Eq.~(\ref{eq13}) and
\begin{equation}
C_{\rm ed}=\frac{1152\gamma_a}{\pi^2}\approx 0.610.
\label{eq47}
\end{equation}
\noindent
In a similar way the gradient of the Casimir force with account
of edge effects is given by
\begin{equation}
\frac{\partial F_{\rm ed}^{\rm IM}(a,0)}{\partial a}=
\frac{\partial F^{\rm IM}(a,0)}{\partial a}
\left(1+\tilde{C}_{\rm ed}\frac{a}{L}\right),
\label{48}
\end{equation}
\noindent
where $\partial F^{\rm IM}(a,0)/{\partial a}$ is defined in
Eq.~(\ref{eq18}) and
\begin{equation}
\tilde{C}_{\rm ed}=\frac{5}{7}C_{\rm ed}\approx 0.436.
\label{eq49}
\end{equation}

We are now in a position to estimate the total relative error
in the Casimir force and its gradient, originating from the
application of the PFA to a cylinder of finite length.
For the Casimir force this is a combination of the
systematic errors
due to the application of the PFA to an infinite cylinder,
$-C_{\rm ex}a/R$, estimated in Sec.~II, and that due to the edge
effects, $C_{\rm ed}a/L$. Taking the most conservative
approach, we consider these systematic errors as random quantities
characterized by a uniform distribution. Then the total error
of the Casimir force determined at a 95\% confidence level
can be obtained from the following statistical rule
\cite{6,65}
\begin{equation}
\delta_{F}^{\rm tot}=\min
\left(C_{\rm ex}\frac{a}{R}+C_{\rm ed}\frac{a}{L},
1.1\sqrt{C_{\rm ex}^2\frac{a^2}{R^2}+
C_{\rm ed}^2\frac{a^2}{L^2}}\right).
\label{eq50}
\end{equation}
\noindent
Taking $R=L=100\,\mu$m we find from Eq.~(\ref{eq50}) that
$\delta_{F}^{\rm tot}\approx 0.07$\% at $a=100\,$nm and
$\delta_{F}^{\rm tot}\approx 0.37$\% at $a=500\,$nm.

To estimate the total error of the gradient of the Casimir
force between a cylinder and a plate, one should use the
exact expression following from Eq.~(\ref{eq14})
\begin{eqnarray}
&&
\frac{\partial F_{\rm ex}^{\rm IM}(a,0)}{\partial a}=
\frac{7\pi^3\hbar cL}{768a^4}\sqrt{\frac{R}{2a}}
\left(1-\tilde{C}_{\rm ex}\frac{a}{L}\right),
\nonumber \\
&&
\tilde{C}_{\rm ex}=\frac{5}{7}C_{\rm ex}\approx 0.2062.
\label{eq51}
\end{eqnarray}
\noindent
This shows that the error in the application of the PFA
to the gradient of the Casimir force between an infinite
cylinder and a plate is less than in the application of
this method to the force. Combining the errors
$\tilde{C}_{\rm ex}a/R$ from Eq.~(\ref{eq51}) and
$\tilde{C}_{\rm ed}a/L$ defined in Eq.~(\ref{eq49})
with the help of the same statistical rule (\ref{eq50}),
one obtains
$\delta_{\partial F/\partial a}^{\rm tot}\approx 0.05$\%
and 0.26\% at $a=100\,$nm and 500\,nm, respectively.

In the dynamic experiments measuring the gradient of the
Casimir force between a sphere and a plate by means of a
micromachined oscillator \cite{30,31,32,33,34} the Au-coated
sphere was attached to the optical fiber. As a result, the
projection of the part of the sphere near the point of attachment
on the plane of an oscillator plate was situated beyond the
plate edge.
As was noted above, in the proposed experiment the cylinder is
attached to an oscillator and the cylinder's projection on the
plate belongs to its interior (see Sec.~VI).
However, in some other experimental configurations it
may happen that the projection of the part of the cylinder
is beyond the edge of the plate.
We now consider the role of edge effects in the case when the
projection of a cylinder axis on the plate is at distance
$L_1<R$ from the edge of a plate [see Fig.~3(b)]. In this case
one half of the cylinder is completely above the plate. As to
 the other half, only the portion $x\leq L_1$ is above
the plate.

We next obtain the expression for the Casimir force
taking into account edge effects,
\begin{equation}
\tilde{F}_{\rm ed}^{\rm IM}(a,0)=\int_{-R}^{L_1}dF_{\rm ed}^{\rm IM}(z,0)
-\frac{2\gamma_a\hbar cL}{(a+H)^3},
\label{eq52}
\end{equation}
\noindent
where $dF_{\rm ed}^{\rm IM}(z,0)$ is defined in Eq.~(\ref{eq45}),
and $H=R-\sqrt{R^2-L_1^2}$  [see Fig.~3(b)].
In the following we assume $L_1,\,H\gg a$.
Here, the upper integration limit $L_1$ [instead of $R$ in
 Eq.~(\ref{eq46})] takes into account that only part of the
 cylinder is above the plate. The second term on the right-hand
 side of Eq.~(\ref{eq52}) is due to the presence of an additional
 real boundary of length $L$.
 This boundary belongs to one of the partial plane plates, with area
 $Ldx$ replacing a cylinder when the PFA is used
 which is situated at a height $a+H$ above the edge of
 an oscillator plate. Equation (\ref{eq52}) can be equivalently
 presented in the form
\begin{equation}
\tilde{F}_{\rm ed}^{\rm IM}(a,0)=\int_{-R}^{R}dF_{\rm ed}^{\rm IM}(z,0)
-\int_{L_1}^{R}dF_{\rm ed}^{\rm IM}(z,0)
-\frac{2\gamma_a\hbar cL}{(a+H)^3}.
\label{eq53}
\end{equation}
\noindent
Performing the integrations and retaining only the leading terms in
the small parameters $a/R$ and $a/H$ we arrive at the result
\begin{eqnarray}
&&
F_{\rm ed}^{\rm IM}(a,0)=
F^{\rm IM}(a,0)\left[
\vphantom{\frac{768\sqrt{2}\gamma_a}{\pi^3}}
1+C_{\rm ed}\frac{a}{L}
\right.
\label{eq54} \\
&&~~~~~
\left.
-
f\left(\frac{L_1}{R}\right)\sqrt{\frac{a}{R}}\frac{a^3}{L_1^3}
+\frac{768\sqrt{2}\gamma_a}{\pi^3}\sqrt{\frac{a}{R}}\frac{a^3}{H^3}
\right].
\nonumber
\end{eqnarray}
\noindent
Here, the following notations are introduced:
\begin{equation}
f\left(\frac{L_1}{R}\right)=\frac{8\sqrt{2}}{525\pi}
\frac{R^4}{L_1^4}f_1\left(\frac{L_1}{R}\right)+
\frac{1536\sqrt{2}\gamma_a}{5\pi^3}
\frac{R^3}{LL_1^2}f_2\left(\frac{L_1}{R}\right),
\label{eq55}
\end{equation}
\noindent
where
\begin{eqnarray}
&&
f_1(z)=120-160z^2+35z^4+13z^7
\label{eq56} \\
&&~~~~~~~~~~~
+4\sqrt{1-z^2}(30-27z^2-z^4-2z^6),
\nonumber \\
&&
f_2(z)=4-5z^2+z^5+\sqrt{1-z^2}(4-3z^2-z^4).
\nonumber
\end{eqnarray}
\noindent
If $L_1=R$ (i.e., the cylinder is completely over
the plate) we have $f_1(1)=f_2(1)=f(1)=0$, and Eq.~(\ref{eq54})
reduces to Eq.~(\ref{eq46}). [Note that in this case the last term
on the right-hand side of Eq.~(\ref{eq54}) is of order $a^3/R^3$
and, thus, should be disregarded in the framework of the PFA.]

It is easily seen that two additional terms
on the right-hand side of Eq.~(\ref{eq54}) are negligibly small.
For example, for the same parameters $R=L=100\,\mu$m,  as above,
and $L_1=25\,\mu$m the contribution of these terms does not exceed
$10^{-4}$\% at $a=100\,$nm and 0.05\%   at $a=500\,$nm.
With the increase of $L_1$ the contribution of the additional
terms in Eq.~(\ref{eq54}) further decreases. For instance, for
$L_1=50\,\mu$m it does not exceed
$5\times 10^{-7}$\% at $a=100\,$nm and
$1.5\times 10^{-4}$\% at $a=500\,$nm.

\section{Proposed experimental setup}\label{exp}

In this section we propose an experiment to measure the Casimir
interaction between a cylinder and a plate using an approach
similar to what was used in previous studies for
sphere-plate geometry \cite{30,31,32,33,34,66}.
 A sensitive micromechanical torsional oscillator (MTO) is at the
heart of the setup since it allows detection of the Casimir force
in a very precise manner. An incorporated optical fiber two-color
interferometer allows the measurement of the fiber-to-substrate
separation at all times. As it was done in our previous
experiments, an electrostatic calibration will allow
us to determine the
interaction force, as well as the separation $a$.

We propose to integrate a metallic cylinder onto a MTO and to
detect the deflections experienced by the micromechanical device
when a metallic plate approaches the cylinder from above (see
Fig.~4 for a schematic of the setup).
Section~II presents theoretical results for the Casimir force
and for the gradient of the Casimir force between a cylinder
and a plate. Both these quantities can be measured using the
suggested setup. The Casimir force is measured in the static
mode \cite{30}. The gradient of the Casimir force is measured
in the dynamic mode, where the separation distance between a
cylinder and a plate is varied harmonically at the resonant
frequency of the MTO \cite{31}.
The experimental setup allows for a force sensitivity of $\sim 0.1\,$pN
and a determination of the resonant frequency of the MTO to better
than 6\,mHz. In the dynamic mode, the gradient of the Casimir force is
obtained from the resonant frequency of the oscillator
\begin{equation}
\omega_{\rm res}=\omega_0\left(1-\frac{b^2}{I\omega_0^2}\frac{\partial
F}{\partial a}\right).
\label{grad}
\end{equation}
\noindent
Here, $\omega_0$ is the natural angular resonance frequency of the
oscillator, $\omega_{\rm res}$ is the angular resonant frequency of
the oscillator in the presence of the Casimir force, $b$ is
the lever arm of the interaction, and $I$ is the moment of
inertia of the oscillator.

In order to minimize errors associated with misalignments between
the torsional axis of the MTO and the cylinder's axis, we propose
to integrate the metallic cylinder onto the MTO using
state-of-the-art semiconducting processing. Currently available
lithographic steppers allow alignment between different structural
layers in a mechanical device with an accuracy of 20\,nm or better.
By using a monolithic fabrication process it would be possible to
directly fabricate a metallic cylinder on top of a MTO similar to
the one previously used. In order to fabricate a cylinder with a
pre-determined cylindrical shape and smooth surface it will be
necessary to use grayscale masks or laser based maskless
lithography technology \cite{optics}. These technologies permit
the creation of three-dimensional structures with smooth surfaces and
precise critical dimensions. A simpler method to fabricate a
cylindrical object on top of a MTO involves deposition and
patterning of photosensitive polymers (photoresist), and subsequent
thermal reflow of the polymer into a curved shape. This technique
relies on surface tension and plastic flow of photoresist and is
used to produce smooth curved objects. Reflow techniques are much
simpler to implement than grayscale lithography techniques, but
they are much more limited regarding the final shape of the
cylindrical objects. The photoresist shape can be transferred to
the MTO using conventional dry etching techniques like reactive
ion etching (RIE). Once the cylindrical shape has been transferred
to the MTO it can be coated with a metallic film by evaporation or
sputtering methods.

The actual cylindrical shape $z=f(x,y)$ can be measured very
precisely, better than 1\,nm error in $z$, by using a non-contact
optical profilometer. Once this shape is known, it can be
introduced in Eq.~(\ref{eq1}) to calculate the Casimir force acting on the
cylinder, as described in Sec.~II.

During the measurements, the cylinder-MTO structure will be
brought close to the Au-coated plane using a 5-axis manipulator
with an angular error of 10$^{-5}$ radians. This angular error
allows for an alignment of the cylinder and the plate to within
the ranges calculated in Sec.~IV.

Further improvements in the alignment between the axis of the MTO
and the plate can be obtained by sacrificing the ability to
perform continuous measurements as a function of separation. In
this approach the oscillator/cylinder unit is manufactured as
previously described, but the Au-coated plate on a Si substrate is
engineered with three equal pillars of known height. The MTO chip
is positioned on top of the pillars and a measurement of the
Casimir interaction at a fixed separation can be performed. By
changing the height of the pillars it is feasible to reproduce a
curve of the Casimir interaction as a function of separation.
Since the height of the pillars can be controlled to better than
10\,nm, misalignments smaller than 1\,$\mu$rad are possible.

\section{Conclusions and discussion}

In the foregoing we have investigated the possibilities of
measuring the thermal Casimir force and its gradient between a
plate and a
microfabricated cylinder attached to a micromachined
oscillator. Using the PFA we have derived the Lifshitz-type
formulas for the thermal Casimir force for a cylinder and a
plate made of real materials. {}From the comparison with exact
results available for an ideal metal cylinder above a plate and
for both ideal and real metals in the configuration of a sphere
above a plate, we have estimated the error resulting from the
use of the PFA. It turns out that for reasonable experimental
parameters this error does not exceed a small fraction of a percent.
In the limiting case of high temperature (large separation) the
analytic expressions for the Casimir force and its gradient are
obtained.

The derived Lifshitz-type formulas were used to perform
numerical computations of the thermal correction to both the
Casimir force and its gradient. In so doing the two alternative
theoretical approaches proposed in the literature (the Drude model
and the plasma model approaches) have been used. The predicted
results of both static and dynamic experiments for the
measurement of the thermal correction were discussed.
Specifically, it was concluded that the thermal correction, as
predicted by the Drude model approach, is large enough to be
measurable at short separations below $1\,\mu$m. With respect to
the thermal correction predicted by the plasma model approach, it
can be measured only at separations of a few micrometers.
The performed computations, when compared with the experimental
data, may shed additional light on the experimental exclusion
of the Drude model approach in Refs.~\cite{30,31,32,33,34}
which was made using only the configuration of a
sphere above a plate.

Keeping in mind the applications of the developed formalism for
comparison with experimental data, we have derived expressions
for the thermal Casimir force and its gradient
taking into account the
nonparallelity of the cylinder and plate. This was done in a
nonmultiplicative manner including the role of correlations between
geometrical and material properties. In the specific case of
ideal metal bodies at zero temperature, the previously obtained
results were confirmed. We have also performed numerical
computations of a correction to the Casimir force
due to nonparallelity of a cylinder and a plate using both
nonmultiplicative and multiplicative approaches, and determined
the application region of the latter.

Taking into account that microfabricated cylinders to be used
in experiments are restricted in length (which may be  of the
order of the
cylinder radius), we have estimated errors in computations
arising from the finiteness of a cylinder. The two
experimentally relevant cases were considered when the
cylinder is completely over the plate or is partially beyond
the plate edge. In both cases it was shown that for typical
experimental parameters the total error due to application of
the PFA and due to edge effects is sufficiently small, and does
not depend on the ratio of the cylinder radius to the cylinder length.
This opens opportunities for using cylinders with
$L\approx R$, with no loss in the accuracy of the theoretical
expressions derived for infinitely long cylinders.

We have presented a scheme of the proposed experiment for
measuring thermal Casimir force and its gradient in the
configuration of an Au-coated plate  and a microfabricated cylinder
attached to the micromachined oscillator. Both the static and
dynamic versions of this experiment are considered. The values
of main experimental parameters that are aimed to achieve
are estimated. The proposed experiment is promising as a
source of additional information on the problem of thermal
Casimir force.

\section*{Acknowledgments}
R.S.D.~acknowledges NSF support through Grant No.\ PHY--0701236
and LANL support through contract No.\ 49423--001--07.
D.L.\ and R.S.D.\ acknowledge support from DARPA grant No.\ 09--Y557.
E.F. was supported in part by DOE under Grant No.~DE-76ER071428.
G.L.K.\ and V.M.M.\ are grateful to the Department of Physics,
Purdue University for financial support.
G.L.K.\ was also partially supported by the Grant of the Russian
Ministry of Education P--184.

\begin{figure*}[h]
\vspace*{-10.cm}
\centerline{\hspace*{2cm}
\includegraphics{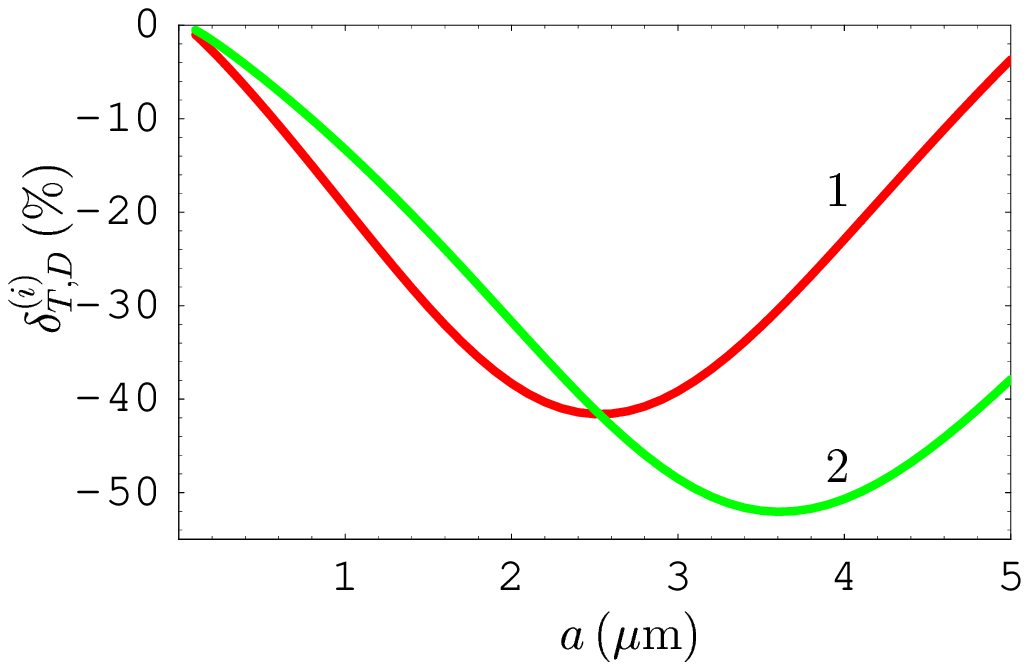}
}
\vspace*{-6.5cm}
\caption{(Color online)
The relative thermal corrections to the Casimir force
and its gradient in cylinder-plate configuration
computed using the Drude model approach at $T=300\,$K
(lines 1 and 2, respectively) are shown as
functions of separation in the region from 0.1 to
$5\,\mu$m.
}
\end{figure*}
\begin{figure*}[h]
\vspace*{-4.cm}
\centerline{\hspace*{-0.1cm}
\includegraphics{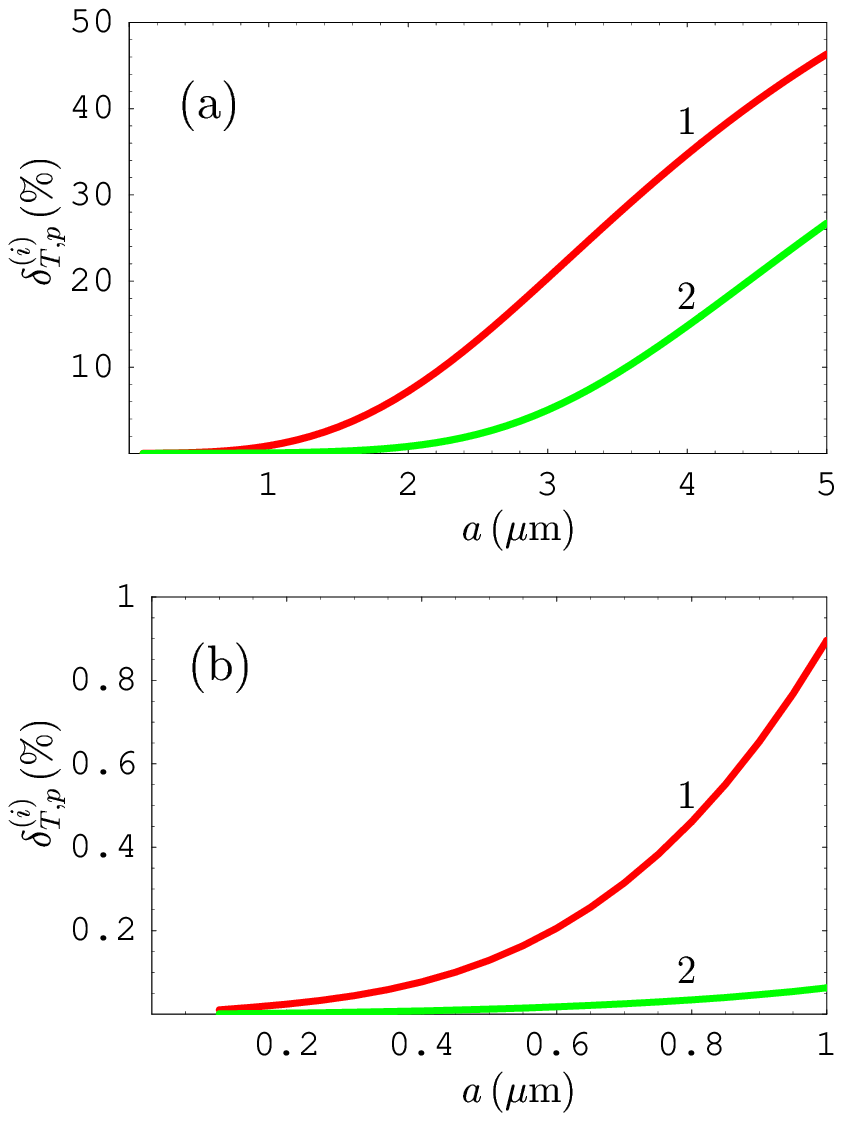}
}
\vspace*{-10.5cm}
\caption{(Color online)
The relative thermal corrections to the Casimir force
and its gradient in cylinder-plate configuration
computed using the plasma model approach at $T=300\,$K
(lines 1 and 2, respectively) are shown as
functions of separation in the region (a) from 0.1 to
$5\,\mu$m and (b) from 0.1 to $1\,\mu$m.
}
\end{figure*}
\begin{figure*}[h]
\vspace*{-0.1cm}
\centerline{
\includegraphics{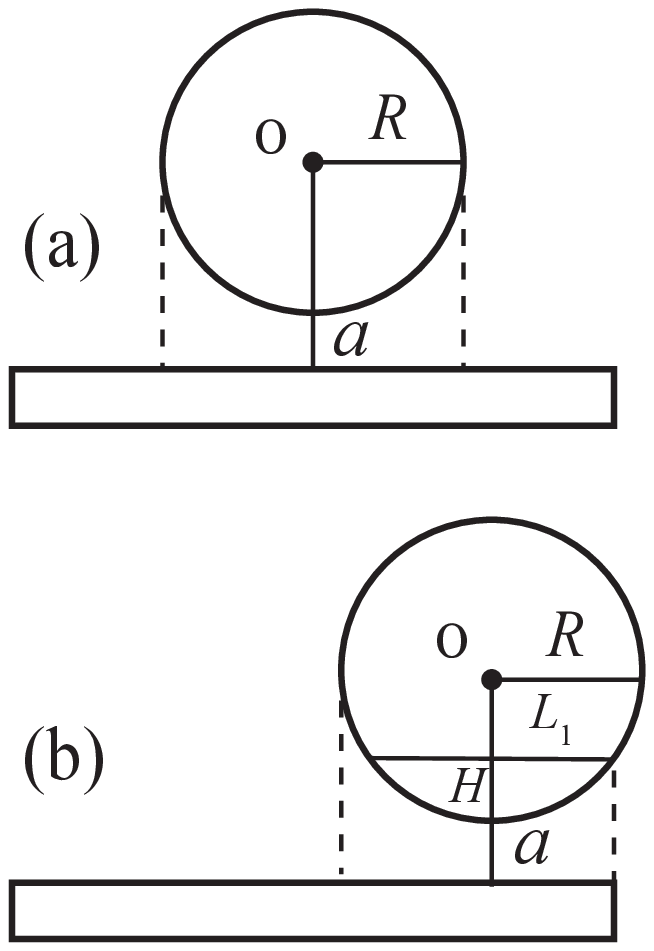}
}
\vspace*{-16.5cm}
\caption{The cylinder and the plate in cases
when (a) the projection of the cylinder on the plane
$(x,y)$ belongs to the interior of the plate and (b)
part of the projection of the cylinder on the plane
$(x,y)$ is beyond the edge of the plate.
}
\end{figure*}
\begin{figure*}[h]
\vspace*{-5.cm}
\centerline{\hspace*{1.5cm}
\includegraphics{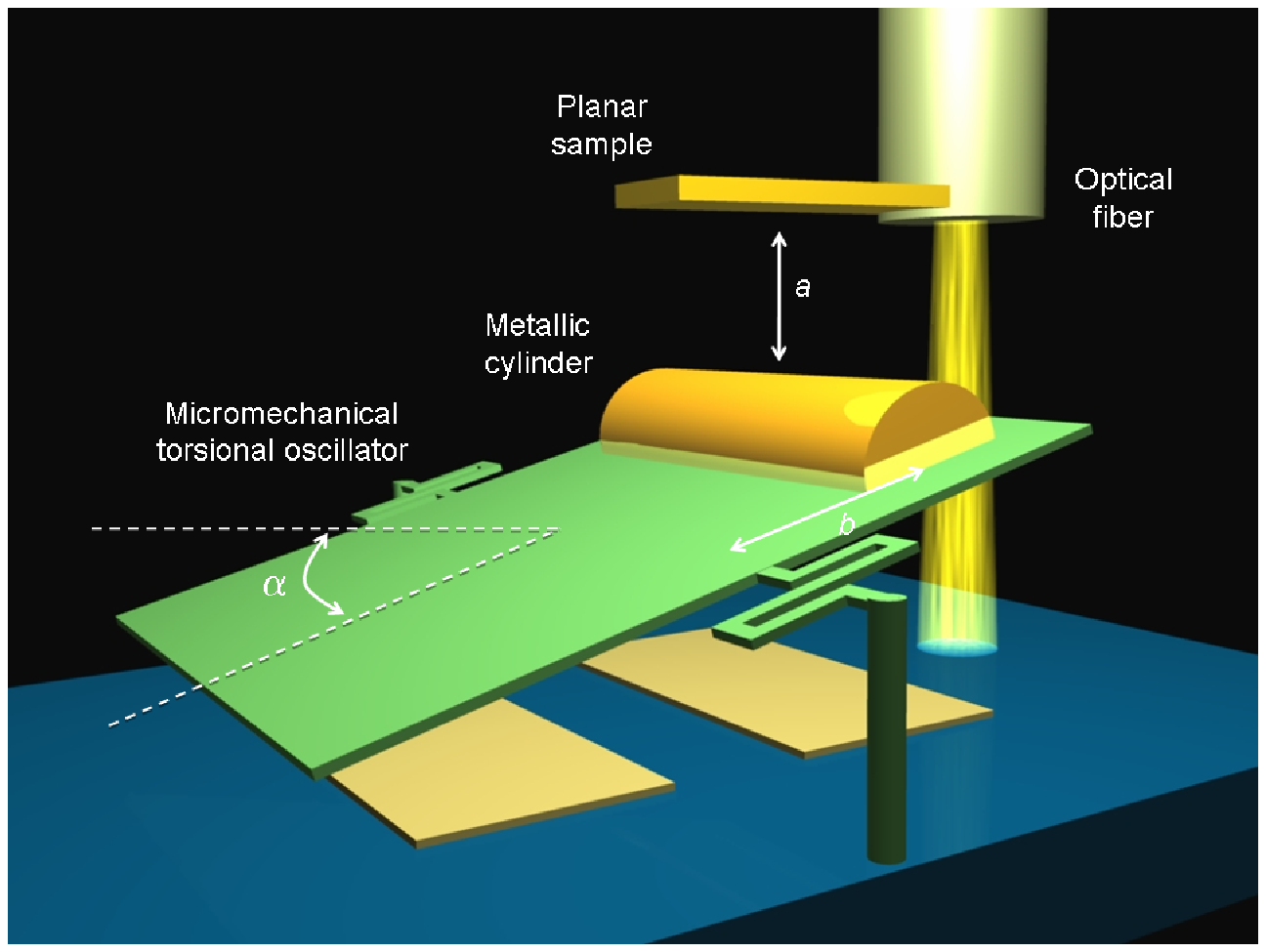}
}
\vspace*{-9.5cm}
\caption{(Color online)
Schematic of the experimental setup (see text for further
discussion). The figure is not to scale.
}
\end{figure*}
\begingroup
\squeezetable
\begin{table}
\caption{\label{tab1}
The lines from 1 to 6 contain the values of the quantity
$\kappa^{(\rm nm)}$ quantifying the effect of the nonzero
angle between a cylinder and a plate on the Casimir force
within the nonmultiplicative approach. In line 7 the
correction factor $\kappa$ taking the nonzero angle in a
multiplicative way is presented. All results are computed
at different values of the parameter
$A_{\theta}=\theta L/(2a)$.}
\begin{ruledtabular}
\begin{tabular}{ccccc}
&$A_{\theta}=0.01$ & $A_{\theta}=0.05$ &$A_{\theta}=0.1$ &
$A_{\theta}=0.5$  \\
\cline{2-5}
$\kappa^{(\rm nm)}(100\,\mbox{nm},A_{\theta})$ &
1.00020 & 1.0051 & 1.0207 & 1.7888 \\
$\kappa^{(\rm nm)}(150\,\mbox{nm},A_{\theta})$ &
1.00021 & 1.0053 & 1.0215 & 1.8230 \\
$\kappa^{(\rm nm)}(200\,\mbox{nm},A_{\theta})$ &
1.00022 & 1.0055 & 1.0221 & 1.8526 \\
$\kappa^{(\rm nm)}(300\,\mbox{nm},A_{\theta})$ &
1.00023 & 1.0057 & 1.0231 & 1.8997 \\
$\kappa^{(\rm nm)}(400\,\mbox{nm},A_{\theta})$ &
1.00023 & 1.0058 & 1.0237 & 1.9335 \\
$\kappa^{(\rm nm)}(500\,\mbox{nm},A_{\theta})$ &
1.00024 & 1.0060 & 1.0242 & 1.9585 \\
$\kappa(A_{\theta})$ &
1.00026 & 1.0066 & 1.0267 & 2.1176
 \\
\end{tabular}
\end{ruledtabular}
\end{table}
\endgroup
\end{document}